\documentclass[
 aip,
 amsmath,amssymb,https://www.overleaf.com/project/625469a90b75a3399fb0d264
]{revtex4-2}

\usepackage{graphicx}
\usepackage{color} 
\usepackage{dcolumn}
\usepackage{bm}

\usepackage[utf8]{inputenc}
\usepackage[T1]{fontenc}
\usepackage{mathptmx}
\usepackage[version=4]{mhchem}
\usepackage{siunitx}
\usepackage{amsfonts}
\usepackage{float}
\usepackage{longtable}
\usepackage{tabularx}
\usepackage{multirow}
\usepackage{rotating}

\usepackage{hyperref}
\hypersetup{
    colorlinks=true,
    linkcolor=blue,
    filecolor=magenta,      
    urlcolor=cyan,
}

\usepackage{xcolor}

\definecolor{neu}{RGB}{0,0,0}
\begin{document}

\preprint{AIP/123-QED}

\title{Cholesky decomposition of two-electron integrals in quantum-chemical calculations with perturbative or finite magnetic fields using gauge-including atomic orbitals}

\author{J{\"u}rgen Gauss}%
 \email{gauss@uni-mainz.de}
\affiliation{Department Chemie, Johannes Gutenberg-Universit{\"a}t Mainz, Duesbergweg 10-14, D-55128 Mainz, Germany}
\author{Simon Blaschke}%
 \email{siblasch@uni-mainz.de}
 \affiliation{Department Chemie, Johannes Gutenberg-Universit{\"a}t Mainz, Duesbergweg 10-14, D-55128 Mainz, Germany}
\author{Sophia Burger}
 \email{soburger@uni-mainz.de}
 \affiliation{Department Chemie, Johannes Gutenberg-Universit{\"a}t Mainz, Duesbergweg 10-14, D-55128 Mainz, Germany}
\author{Tommaso Nottoli}%
 \email{tommaso.nottoli@phd.unipi.it}
\affiliation{Dipartimento di Chimica e Chimica Industriale, Universit\`{a} di Pisa, Via G. Moruzzi 13, I-56124 Pisa, Italy}
\author{Filippo Lipparini}%
 \email{filippo.lipparini@unipi.it}
\affiliation{Dipartimento di Chimica e Chimica Industriale, Universit\`{a} di Pisa, Via G. Moruzzi 13, I-56124 Pisa, Italy}
\author{Stella Stopkowicz}%
 \email{stella.stopkowicz@uni-mainz.de}
\affiliation{Department Chemie, Johannes Gutenberg-Universit{\"a}t Mainz, Duesbergweg 10-14, D-55128 Mainz, Germany}
\affiliation{Fachrichtung Chemie, Universit{\"a}t des Saarlandes, Campus B2.2, D-66123 Saarbr{\"u}cken, Germany }
\
%
\date{\today}

\begin{abstract}
A rigorous analysis is carried out concerning the use of Cholesky decomposition (CD) of two-electron integrals in the case of quantum-chemical calculations with finite or perturbative magnetic fields and gauge-including atomic orbitals. We investigate in particular how permutational symmetry can be accounted for in such calculations and how this symmetry can be exploited to reduce the computational requirements. A modified CD procedure is suggested for the finite-field case that roughly halves the memory demands for the storage of the Cholesky vectors. The resulting symmetry of the Cholesky 
vectors also enables savings in the computational costs. 
For the derivative two-electron integrals in case of a perturbative magnetic field we derive CD expressions by means of a first-order Taylor 
expansion of the corresponding finite magnetic-field formulas with the field-free case as reference point. The perturbed Cholesky vectors are shown to be antisymmetric (as already proposed by Burger {\sl et al.} ({\sl J. Chem. Phys.}, {\bf 155}, 074105 (2021))) and the corresponding expressions enable significant savings in the required integral evaluations (by a factor of about four) as well as in the actual construction of the Cholesky vectors (by means of a two-step procedure similar to the one presented by Folkestad {\sl et al.} ({\sl J. Chem. Phys.}, {\bf 150}, 194112 (2019)) and Zhang {\sl et al.} ({\sl J. Phys. Chem. A}, {\bf 125}, 4258-4265 (2021))). Numerical examples with cases involving several hundred basis functions verify our suggestions concerning CD in case of finite and perturbative magnetic fields. 
\end{abstract}
\maketitle

\section{\label{intro}{Introduction}}

The use of Cholesky decomposition (CD) for a compact representation of the two-electron integrals in quantum-chemical calculations has attracted a lot interest
during the last years (for a recent review, see Ref.~\onlinecite{Aquilante11}). 
Beebe and Linderberg \cite{Beebe77} were the first to suggest the use of CD in quantum-chemical calculations and Koch and coworkers \cite{Koch03} later demonstrated in a convincing manner that the CD of the two-electron integrals can significantly reduce the computational cost in large-scale quantum-chemical computations. Since then many quantum-chemical schemes have been combined with CD; to be mentioned are implementations using CD for Hartree-Fock (HF) calculations,\cite{Koch03,Nottoli21b} second-order M{\o}ller-Plesset (MP2) perturbation theory,\cite{Koch03,Aquilante07b} complete-active space self-consistent-field (CASSCF) treatments,\cite{Aquilante08a,Nottoli21} multiconfigurational second-order perturbation theory (CASPT2),\cite{Aquilante08b} and coupled-cluster (CC) and equation-of-motion coupled-cluster (EOM-CC) approaches.\cite{Epifanovsky13} CD has not only been used in energy calculations and corresponding implementations for geometrical gradients\cite{Delcey14,Feng19,Schnack22}but also for the computation of NMR shieldings\cite{Burger21,Nottoli22b} together with the use of gauge-including atomic orbitals (GIAOs).\cite{London37,Ditchfield72,Hameka58,Wolinski90,Helgaker88} Furthermore,
the efficiency of CD for finite magnetic-field quantum-chemical calculations has been demonstrated.\cite{Blaschke22} 

Despite the tremendous progress within recent years concerning CD, there still is a need and room for further computational improvements. To speed up CD-based quantum-chemical calculations, Aquilante, Lindh, and Pedersen\cite{Aquilante07a,Pedersen09} suggested a one-center CD scheme in the spirit of density fitting.\cite{Whitten73,Dunlap79,Eichkorn95} Indeed, an excellent discussion of the relationship between CD and density fitting can be found in Ref.~\onlinecite{Pedersen09} where a so-called atomic CD scheme has been introduced together with the very useful notion of a Cholesky basis. It should be noted that the atomic CD and the one-center CD variants come with
additional approximations and thus reduced accuracy. Another promising suggestion is the use of
method-specific CD schemes\cite{Boman08} in which the CD is tailored towards the needs of the actual computation. Most recently, Folkestad {\it et al.}\cite{Folkestad19} presented an efficient two-step CD algorithm in which in a first step the Cholesky basis is determined and only in a second step the actual Cholesky vectors are constructed. This two-step CD procedure has been shown to be highly efficient, thus allowing computations with up to 80000 basis functions.\cite{Folkestad19} A further refinement of the two-step procedure by Folkestad {\it et al.} has been proposed by Zhang {\it et al.}\cite{Zhang21}

An issue which is in particular puzzling in case of CD based quantum-chemical computations in the presence of finite or perturbative magnetic fields is the role of permutational symmetry. As it is well known, the usual two-electron integrals possess eightfold permutational symmetry which can be handled within the CD step in a straightforward manner by simply restricting the range of indices for the basis-function pairs. However, in case of magnetic fields this symmetry is reduced\cite{Stopkowicz15} (either for the two-electron integrals themselves in case of a finite magnetic field or the corresponding derivative integrals in case of a perturbative
magnetic field) and it is less clear how one can account for this in the CD step. In the initial CD implementation for finite magnetic fields,\cite{Blaschke22} symmetries of the Cholesky vectors were not exploited. For the treatment of derivative integrals in case of perturbative magnetic fields,\cite{Burger21} antisymmetry of the perturbed Cholesky vectors was deduced and exploited in the actual implementation. However, a more rigorous discussion seems to be warranted together with an exploration of further possible savings in the corresponding CD procedure for the magnetic derivative two-electron integrals.

In this paper, we will analyze the role of permutational symmetry when using CD in quantum-chemical calculations with a finite or perturbative magnetic field. In particular, we devise an improved CD procedure for use with finite magnetic-field calculations.
We start with a brief recapitulation of CD in quantum-chemical calculations, before we 
analyze the consequences of the permutational symmetry of the two-electron integrals in CD based calculations with a finite magnetic field. Based on these findings we propose a modified CD scheme which explicitly takes permutational symmetry into account and has the potential for significant computational savings. The discussion is followed by a corresponding analysis for the two-electron integral derivatives that appear in calculations with perturbative magnetic fields and that are handled by differentiation of the original (unperturbed) CD procedure.\cite{Burger21} We will discuss the role of permutational symmetry, justify the ad-hoc assumption of antisymmetry of the perturbed Cholesky vectors used in Ref.\onlinecite{Burger21} and propose further improvements including the use of a two-step CD procedure to generate the perturbed Cholesky vectors. Finally, we provide numerical evidence for the suggested improvements by showing examples with several hundred basis functions and by comparing the required computational resources of our improved schemes to those of the original procedures.

\section{Theory}

\subsection{\label{sec2a}{Cholesky decomposition of two-electron integrals}}

As suggested by Beebe and Linderberg,\cite{Beebe77,Koch03} a compact representation of the two-electron integrals
$(\sigma \rho | \nu \mu)$ required in quantum-chemical calculations can be obtained via a Cholesky decomposition (CD):
\begin{eqnarray}
\label{cd1}
(\sigma \rho | \nu \mu) \approx \sum_{P=1}^{N_\text{rank}} L_{\sigma \rho}^P (L_{\mu\nu}^P)^*,
\end{eqnarray}
i.e., a decomposition technique that can be applied to any positive semi-definite Hermitian matrix. In Eq.~(\ref{cd1}), $N_\text{rank}$ denotes the rank of the decomposition, $L_{\mu \nu}^P$ are the components of the $P$-th Cholesky vector (CV), and Greek indices $\mu, \nu, \dots$ represent the atomic-orbital basis functions $\chi_{\mu}, \chi_{\nu}, \dots$. The CVs themselves can be determined in an iterative procedure via
\begin{eqnarray}
\label{cd2}
L_{\sigma \rho}^P = \widetilde{(\mu \nu | \nu \mu)}^{- \frac{1}{2}}
\left\{(\sigma \rho | \nu \mu ) - \sum_{R=1}^{P-1} L_{\sigma \rho}^R {L_{\mu\nu}^R}^*\right\}
\end{eqnarray} 
where a new CV, corresponding to the index pair $\mu$ and $\nu$, is chosen by means of a (partial) pivoting procedure.\cite{Koch03} The updated
diagonal elements of the two-electron integral matrix are given by
\begin{eqnarray}
\widetilde{(\mu \nu | \nu \mu)} = (\mu \nu | \nu\mu) - \sum_{R=1}^{P-1} L_{\mu \nu}^R (L_{\mu \nu}^R)^*.
\end{eqnarray}
Iterations are continued and new vectors are added until the largest updated diagonal element is smaller than a predefined Cholesky threshold $10^{-\delta}$. 
This threshold also sets the accuracy of the decomposition, as the Cauchy-Schwarz inequality\cite{Koch03} ensures that the error of the approximated two-electron integrals is in absolute terms always smaller than $10^{-\delta}$.

In the two-step algorithm proposed by Folkestad {et al.}\cite{Folkestad19} the first step involves the set up of the Cholesky basis $\{P\}$ which can be done by selecting products of the basis function (in the following referred to as Cholesky basis functions (CBFs)) from suitable subsets of the full product basis ${\cal B} = \{\chi_\mu \chi_\nu, \mu \ge \nu\}$. The second step then consists of computing the CVs
in the unnormalized Cholesky basis (note that we assume here the Coulomb metric)
\begin{eqnarray}
{\widetilde L}_{\sigma\rho}^P = (\sigma \rho| P)\end{eqnarray}
followed by a transformation of the vectors into the orthonormal Cholesky basis
\begin{eqnarray}
L_{\sigma\rho}^P = \sum_Q {\widetilde L}_{\sigma\rho}^Q ( Q|P )^{-1/2},
\end{eqnarray}
where we denote by $(Q|P)^{1/2}$ the Cholesky decomposition of the metric.

While both the original and the two-step algorithm exhibit the same formal scaling for the formation of the CVs from the computed integrals, i.e. $N_\text{CD}^2 N_\text{bf}^2$
with $N_\text{CD}$ as the number of CVs and $N_\text{bf}$ as the number of basis functions, 
the advantage of the two-step procedure is that this transformation can be carried out using efficient linear algebra routines, and in particular by first computing the CD of the metric and then solving the linear system
\begin{equation}
    \label{eq:OrthoCD}
    \sum_Q (P|Q)^{1/2}L^Q_{\sigma\rho} = (\sigma\rho|P).
\end{equation}
Clearly, this is much more efficient than the Gram-Schmidt like transformation of the original procedure that consists in sequences of scalar products with the already available vectors.

\subsection{Cholesky decomposition and permutational symmetry} 
An issue that needs to be considered in the application of CD to the two-electron integrals
is permutational symmetry. In calculations without magnetic fields, the corresponding two-electron integrals exhibit eightfold permutational symmetry:
\begin{eqnarray}
(\sigma \rho | \mu \nu ) = 
(\rho \sigma  | \mu \nu ) =
(\sigma \rho | \nu \mu ) =
(\rho \sigma | \nu \mu ) =
(\mu \nu | \sigma \rho ) =
(\mu \nu | \rho \sigma ) =
(\nu \mu | \sigma \rho ) =
(\nu \mu | \rho \sigma ). \nonumber \\
\end{eqnarray}
This permutational symmetry can be easily accounted for by actually
decomposing the restricted two-electron integral matrix
\begin{eqnarray}
I_{\sigma \rho, \mu \nu} = (\sigma \rho | \nu \mu)   {\rm \ \ with\ \  } \sigma \ge \rho; \mu\ge \nu
\end{eqnarray}
which also implies that the CVs are symmetric with respect to an interchange of the two indices $\sigma$ and $\rho$:
\begin{eqnarray}
L_{\sigma \rho}^P = L_{\rho \sigma}^P.
\end{eqnarray}

However, the situation is more difficult in case of finite magnetic-field calculations using GIAOs,\cite{Tellgren08,Stopkowicz15,Stopkowicz18} as in this case the complex two-electron integrals only exhibit fourfold permutational symmetry: 
\begin{eqnarray}
\label{perm2}
(\sigma \rho | \nu \mu ) = 
(\rho \sigma | \mu \nu )^\ast =
(\nu \mu | \sigma \rho ) =
(\mu \nu | \rho \sigma )^\ast,
\end{eqnarray}
which cannot be accounted for by restricting the ranges of the two indices 
for electron 1 and 2, respectively. 

From the symmetry relations in Eq.~(\ref{perm2}), it follows that, with the CVs in case of a finite magnetic field calculation denoted as $M_{\sigma \rho}^P$,
\begin{eqnarray}
\label{perm3}
(\sigma \rho | \nu \mu) & \approx& \sum_P M
_{\sigma \rho}^P ({M_{\mu \nu}^P})^\ast\nonumber \\
&\approx& \sum_P ({M
_{\rho \sigma}^P})^\ast {M_{\nu \mu}^P}.
\end{eqnarray}
However, it would be too hasty to deduce from 
these relations Hermitian symmetry for the CVs in case of a finite magnetic-field calculation. Actually, we have to consider the fact that application of permutational symmetry does not map each CBF/CV on itself, rather that one might get a mapping from a CBF/CV for $P \rightarrow (\mu\nu)$ to a CBF/CV for $P^\prime \rightarrow (\nu \mu)$. We thus conclude that the CD in case of a finite magnetic-field calculation should lead (apart for products of basis functions that are real) to pairs of CVs, with indices $P\rightarrow (\mu\nu)$ and $P^\prime\rightarrow (\nu\mu)$, that are conjugate to each other.
From the permutational symmetry relation in Eq.~(\ref{perm2}) one can then deduce that the following should hold:
\begin{eqnarray}
\label{symmagcv}
M_{\sigma \rho}^{P\rightarrow (\mu \nu)} = 
(M_{\rho \sigma}^{P^\prime\rightarrow (\nu \mu)})^\ast.
\end{eqnarray}
This symmetry relation has not been used in Ref. \onlinecite{Blaschke22}, as it is not compatible with the CD procedure set up there. However, the exploitation of the symmetry in Eq.\eqref{symmagcv} should enable us to deal with the unsolved issues concerning the use of CD in finite magnetic-field calculations which are  (1) the question how to choose in the pivoting procedure between indices $P \rightarrow (\mu\nu)$ and $P^\prime \rightarrow (\nu \mu)$, which both belong to diagonal elements of same magnitude, (2) the pending issue how to preserve symmetry between the pairs $(\mu \nu)$ and $(\nu \mu)$ in the CD, and (3) computational efficiency. In the next section we describe a modified CD procedure that takes these considerations into account.

\subsection{Modified Cholesky decomposition procedure for finite magnetic-field calculations using GIAOs} 

The CD procedure proposed in the following ensures a treatment of conjugate pairs $P \rightarrow (\mu\nu)$ and $P^\prime \rightarrow (\nu \mu)$ on an equal footing, while otherwise it closely follows the original CD procedure. This means that the selection of a new (pair of) CBF(s)/CV(s) is done by means of (partial) pivoting, i.e., a new (pair of) CBF(s)/CV(s) is selected by identifying the largest updated diagonal element of the two-electron integral matrix. To be more specific, our modified CD procedure consists of the following steps:
\begin{enumerate}
    \item compute the diagonal elements $(\mu \nu | \nu \mu)$ for all $\chi_{\mu}\chi_{\nu} \in {\cal B}$;
    \item determine the indices $\mu$ and $\nu$ of the largest (updated) diagonal element (see Eqs.~(\ref{updated1}) and (\ref{updated2}) for the update of the diagonal elements)
    \begin{eqnarray}
    P \leftarrow (\mu, \nu) \in {\cal B} {\rm\ with\ }
    |\widetilde{(\mu \nu | \nu \mu)}| \ge |\widetilde{(\sigma \rho| \rho \sigma)}| {\rm\ for\ all\ } (\sigma, \rho) \in {\cal B};
    \end{eqnarray}
    \item in case of a real product $\chi_{\mu} \chi_{\nu}$, i.e., a product for which the resulting phase factor is independent of the magnetic field $\bf B$ and just one, proceed as usual, i.e., add the new product function with index $P$ to the Cholesky basis and compute the corresponding CV in the unorthogonalized Cholesky basis\begin{eqnarray}
    {\widetilde M_{\sigma \rho}}^{P \rightarrow (\mu \nu)} = ( \sigma \rho| \nu \mu),
    \end{eqnarray} transform the vector (in a Gram-Schmidt like procedure) into the orthogonalized Cholesky basis
    \begin{eqnarray}
    \label{cdeq1}
    \overline {M}_{\sigma \rho}^{P \rightarrow (\mu \nu)}  = {\widetilde M_{\sigma \rho}}^{P \rightarrow (\mu \nu)} - \sum_Q^{P-1} M_{\sigma \rho}^Q (M_{\mu \nu}^Q)^\ast,
    \end{eqnarray}
    and "normalize" the vector
    \begin{eqnarray}
    M_{\sigma \rho}^{P \rightarrow (\mu \nu)} = 
    \overline {M}_{\sigma \rho}^{P \rightarrow (\mu \nu)}/ \sqrt{\widetilde {( \mu \nu | \nu \mu)}};
    \end{eqnarray}
    \item in case of a complex product $\chi_{\mu} \chi_{\nu}$, pick two new CBFs
    with indices $P \rightarrow (\mu\nu)$ as well as $P^\prime \rightarrow (\nu \mu)$ and compute the corresponding CVs in the unorthogonalized representation
    \begin{eqnarray}
    {\widetilde M_{\sigma \rho}}^{P \rightarrow (\mu \nu)} = ( \sigma \rho| \nu \mu) \\
    {\widetilde M_{\sigma \rho}}^{P^\prime \rightarrow (\nu \mu)} = ( \sigma \rho| \mu \nu),
    \end{eqnarray} 
    transform both CVs (in a Gram-Schmidt like procedure) into a partially orthogonal representation (i.e., one in which the two new product functions have been orthogonalized with respect to all previous CBFs but not with respect to each other; the sum  (as well as the sum in Eq.~(\ref{cdeq1})) thus runs over all CVs up to $P-1$, thereby assuming that $P^\prime=P+1$ and that the sum also includes the conjugate CVs)
    \begin{eqnarray}
     \overline {M}_{\sigma \rho}^{P \rightarrow (\mu \nu)}  = {\widetilde M_{\sigma \rho}}^{P \rightarrow (\mu \nu)} - \sum_{Q=1}^{P-1} M_{\sigma \rho}^Q (M_{\mu \nu}^Q)^\ast, \\
      \overline {M}_{\sigma \rho}^{P^\prime \rightarrow (\mu \nu)}  = {\widetilde M_{\sigma \rho}}^{P^\prime \rightarrow (\nu \mu)} - \sum_{Q=1}^{P-1} M_{\sigma \rho}^Q (M_{\nu \mu}^Q)^\ast,
    \end{eqnarray}and then carry out the transformation into a fully orthonormalized basis (based on a symmetric orthonormalization of the two CBFs with indices $P$ and $P^\prime$) 
    \begin{eqnarray}
    \label{symortho}
    M_{\sigma \rho}^{P \rightarrow (\mu \nu)} = 
    (S^{-1/2})_{PP} {\overline M}_{\sigma \rho}^{P \rightarrow (\mu \nu)} + 
    (S^{-1/2})_{PP^\prime}{\overline M}_{\sigma \rho}^{P^\prime \rightarrow (\nu \mu)}\\
    M_{\sigma \rho}^{P^\prime \rightarrow (\nu \mu)} = 
    (S^{-1/2})_{P^\prime P} {\overline M}_{\sigma \rho}^{P \rightarrow (\mu \nu)} +
    (S^{-1/2})_{P^\prime P^\prime}{\overline M}_{\sigma \rho}^{P^\prime \rightarrow (\nu \mu)},\end{eqnarray}
    where the matrix $S$ is the overlap matrix (using the Coulomb metric) of the two CBFs with indices $P$ and $P^\prime$ (after orthogonalization against the previous CBFs)
    \begin{eqnarray}
    S_{PP} = \overline {M}_{\mu \nu}^{P \rightarrow (\mu \nu)},\\
    S_{PP^\prime} = \overline {M}_{\mu \nu}^{P^\prime \rightarrow (\nu \mu)},\\
    S_{P^\prime P} = \overline {M}_{\nu \mu}^{P \rightarrow (\mu \nu)},\\
    S_{P^\prime P^\prime} = \overline {M}_{\nu \mu}^{P^\prime \rightarrow (\nu \mu)};
    \end{eqnarray}
    \item update the remaining diagonal elements  \begin{eqnarray}
    \label{updated1}
    \widetilde {(\sigma \rho | \rho \sigma)} &\Leftarrow&
    \widetilde {(\sigma \rho | \rho \sigma)} - M_{\sigma \rho}^{P \rightarrow (\mu \nu)}
    (M_{\sigma \rho}^{P \rightarrow (\mu \nu)})^\ast \qquad\qquad\qquad\qquad\qquad\qquad (P \rightarrow {\rm  real\ product)} \nonumber \\ \\    \label{updated2}\widetilde {(\sigma \rho | \rho \sigma)} &\Leftarrow&
    \widetilde {(\sigma \rho | \rho \sigma)} - M_{\sigma \rho}^{P \rightarrow (\mu \nu)}
    (M_{\sigma \rho}^{P \rightarrow (\mu \nu)})^\ast  - M_{\sigma \rho}^{P^\prime \rightarrow (\nu \mu)}
    (M_{\sigma \rho}^{P^\prime \rightarrow (\nu \mu)})^\ast \qquad (P \rightarrow {\rm complex\ product)}\nonumber \\\end{eqnarray}and if the largest updated diagonal element
   is still above the predefined Cholesky threshold $10^{-\delta}$, return to step 2.
\end{enumerate}
It is easily seen that the CVs determined in this way obey the symmetry relation given in Eq.~(\ref{perm3}). This means that savings in the storage requirements for the CVs by about a factor of two are possible, as in case of real products the CVs constitute Hermitian matrices with respect to the two indices $\sigma$ and $\rho$, while in the complex case only one of the two conjugate CVs need to be stored.

Note that the limiting case for ${\bf B} \rightarrow {\bf 0}$ is well defined for indices $P$ that correspond to real products
\begin{eqnarray}
\label{limit1}
M^{P\rightarrow (\mu\nu)}_{\sigma\rho} \rightarrow L_{\sigma \rho}^P,\qquad\qquad\qquad\qquad\qquad\qquad (P \rightarrow {\rm  real\ product)}
\end{eqnarray} while for complex products both CVs, i.e., the one with index $P$ and the one with index $P^\prime$, collapse 
\begin{eqnarray}
\label{limit2}
M_{\sigma \rho}^{P\rightarrow (\mu\nu)} \rightarrow \frac{1}{\sqrt{2\widetilde{(\mu\nu|\nu\mu)}}}\overline{M}^{P\rightarrow (\mu\nu)}_{\sigma\rho}\rightarrow \frac{1}{\sqrt{2}}L^P_{\sigma\rho}\phantom{\quad\qquad\qquad\qquad\qquad (P \rightarrow {\rm  complex\ product)}} \nonumber \\
M_{\sigma \rho}^{P^\prime \rightarrow (\nu\mu)} \rightarrow \frac{1}{\sqrt{2\widetilde{(\mu\nu|\nu\mu)}}}\overline{M}^{P^\prime\rightarrow (\nu\mu)}_{\sigma\rho}\rightarrow \frac{1}{\sqrt{2}}L^P_{\sigma\rho}\quad\qquad\qquad\qquad\qquad (P \rightarrow {\rm  complex\ product)}\nonumber \\
\end{eqnarray}
and one just considers the CV with index $P$.
Eq.~(\ref{limit2}) implies that the field-free limit of the inverse square root of the overlap matrix needed for 
the orthonormalization of the Cholesky basis in Eq.~(\ref{symortho}) is just the unit matrix scaled by $(2\widetilde{(\mu\nu|\nu\mu)})^{-1/2}$.

\subsection{\label{sec2c}Cholesky decomposition of magnetic two-electron integral derivatives}

For the magnetic derivatives of the two-electron integrals, a CD procedure can be set up by differentiating the equations of the original CD procedure as, for example, described in Ref.~\onlinecite{Burger21}. A complication that arises there is, as already mentioned, the handling of permutational symmetry, as the original eightfold symmetry of the two-electron integrals is lost when switching on a magnetic field. We derive in the following expressions for these derivative two-electron 
integrals via first-order Taylor expansions of the CD expressions given for the two-electron integrals in case of a finite magnetic field.

A Taylor expansion of the CD for the two-electron integrals in a finite magnetic field $\bf B$ up to first order around ${\bf B} ={\bf 0}$ yields:
\begin{eqnarray}
\label{eq:taylor}
(\sigma \rho | \nu \mu)_{\bf B} &=& (\sigma \rho| \nu \mu)_{{\bf B} = {\bf 0}} + \left(\frac{\partial (\sigma \rho| \nu \mu)}{\partial {\bf B}}\right)_{{\bf B} = {\bf 0}} {\bf B} + \dots \\
&=&
\sum_P M_{\sigma \rho}^P (M_{\mu \nu}^P)^\ast  + \sum_{P \rightarrow{\rm real\ product}} \left\{M_{\sigma \rho}^P \left(\frac{\partial M_{\mu \nu}^P}{\partial \bf B}\right)^\ast_{{\bf B} = {\bf 0}} + \left(\frac{\partial M_{\sigma \rho}^P}{\partial \bf B}\right)_{{\bf B} = {\bf 0}} (M_{\mu \nu}^P)^\ast  \right\} {\bf B} \nonumber \\ && + \sum_{P \rightarrow{\rm complex\ product}} \left\{M_{\sigma \rho}^P \left(\frac{\partial M_{\mu \nu}^P}{\partial \bf B}\right)^\ast_{{\bf B} = {\bf 0}} + \left(\frac{\partial M_{\sigma \rho}^P}{\partial \bf B}\right)_{{\bf B} = {\bf 0}} (M_{\mu \nu}^P)^\ast \right. \nonumber \\ && \left. \qquad\qquad\qquad\qquad +
M_{\sigma \rho}^{P^\prime} \left(\frac{\partial M_{\mu \nu}^{P^\prime}}{\partial \bf B}\right)^\ast_{{\bf B} = {\bf 0}} 
+ \left(\frac{\partial M_{\sigma \rho}^{P^\prime}}{\partial \bf B}\right)_{{\bf B} = {\bf 0}} (M_{\mu \nu}^{P^\prime})^\ast\right\} {\bf B} + \dots,
\end{eqnarray}
where $M_{\sigma \rho}^P$ denotes the corresponding CVs in the limit ${\bf B} \rightarrow {\bf 0}$. As the first derivatives of the two-electron integrals with respect to the magnetic-field components are purely imaginary, it follows that the perturbed CVs must be also purely imaginary as well and, furthermore, from Eq.~(\ref{perm2}) it can be deduced that
\begin{eqnarray}
\label{psym1}
 \left(\frac{\partial M_{\sigma \rho}^P}{\partial \bf B}\right)_{{\bf B} = {\bf 0}} = -  \left(\frac{\partial M_{\rho \sigma}^P}{\partial \bf B}\right)_{{\bf B} = {\bf 0}}
 \qquad\qquad\qquad( P \rightarrow {\rm real\ product}) \\
 \label{psym2}
 \left(\frac{\partial M_{\sigma \rho}^P}{\partial \bf B}\right)_{{\bf B} = {\bf 0}} = -  \left(\frac{\partial M_{\rho \sigma}^{P^\prime}}{\partial \bf B}\right)_{{\bf B} = {\bf 0}}.
 \qquad\qquad\qquad\qquad( P \rightarrow {\rm complex\ product}) 
\end{eqnarray}
To proceed with the evaluation of the integral derivatives from Eq.~(\ref{eq:taylor}), the zero-field limits of the perturbed CVs are needed. These can be obtained from Eqs.~(\ref{limit1}) and (\ref{limit2}), by using the limit of the overlap matrix used to orthogonalize the pairs of vectors stemming from a complex product, giving
\begin{eqnarray}
\label{eq:derlim1}
\left(\frac{\partial M_{\sigma \rho}^P}{\partial \bf B}\right)_{{\bf B} = {\bf 0}} = \left (\frac{\partial L_{\sigma\rho}^P}{\partial \bf B}\right )_{{\bf B} = {\bf 0}}
\qquad\qquad\qquad (P \rightarrow\rm  real\ product)
\end{eqnarray}
and
\begin{eqnarray}
\label{eq:derlim2}
\left(\frac{\partial M_{\sigma \rho}^{P\rightarrow(\mu\nu)}}{\partial \bf B}\right)_{{\bf B} = {\bf 0}} \rightarrow \frac{1}{\sqrt{2\widetilde{(\mu\nu|\nu\mu)}}}\left(\frac{\partial \overline{M}_{\sigma \rho}^{P\rightarrow(\mu\nu)}}{\partial \bf B}\right)_{{\bf B} =\bf 0} \rightarrow
\frac{1}{\sqrt{2}}\left (\frac{\partial L_{\sigma\rho}^P}{\partial \bf B}\right )_{{\bf B} = {\bf 0}} \nonumber \\
\left(\frac{\partial M_{\sigma \rho}^{P^\prime\rightarrow(\nu\mu)}}{\partial \bf B}\right)_{{\bf B} = {\bf 0}} \rightarrow\frac{1}{\sqrt{2\widetilde{(\mu\nu|\nu\mu)}}}\left(\frac{\partial \overline{M}_{\sigma \rho}^{P^\prime\rightarrow(\nu\mu)}}{\partial \bf B}\right)_{{\bf B} =\bf 0} \rightarrow  \frac{1}{\sqrt{2}}\left (\frac{\partial L_{\sigma\rho}^P}{\partial \bf B}\right )_{{\bf B} = {\bf 0}}. 
\nonumber \\ \qquad\qquad\qquad (P \rightarrow\rm  complex\ product)
\end{eqnarray}
Note that this also allows us to define a symmetrized limit expression for the complex product case which will be useful in later derivations:
\begin{eqnarray}
\label{pertlc}
 \frac{1}{\sqrt 2} \left\{ \left(\frac{\partial M_{\sigma \rho}^P}{\partial \bf B}\right)_{{\bf B}= {\bf 0}} + \left(\frac{\partial M_{\sigma \rho}^{P^\prime}}{\partial \bf B}\right)_{{\bf B}= {\bf 0}} \right\} \rightarrow \left(\frac{\partial L_{\sigma \rho}^P}{\partial \bf B}\right)_{{\bf B}= {\bf 0}} 
\end{eqnarray}
By substituting Eqs.~(\ref{limit1}), (\ref{limit2}), (\ref{eq:derlim1}), and (\ref{eq:derlim2}) into Eq.~(\ref{eq:taylor}), and by exploiting the relations in Eqs.~(\ref{psym1}) and (\ref{psym2}), we get
\begin{eqnarray}
\label{intder}
\left(\frac{\partial (\sigma \rho| \nu \mu)}{\partial {\bf B}}\right)_{{\bf B} = {\bf 0}}  =
\sum_P \left\{
\left(\frac{\partial L_{\sigma \rho}^P}{\partial \bf B}\right)_{{\bf B} = {\bf 0}} L_{\mu \nu}^P - L_{\sigma \rho}^P \left(\frac{\partial L_{\mu \nu}^P}{\partial \bf B}\right)_{{\bf B} = {\bf 0}} \right\}.
\end{eqnarray}
Note that the sum in Eq.~(\ref{intder}) solely runs over vectors with indices $P$ and that there is no need to include the conjugate vectors with indices $P^\prime$. 
We can furthermore deduce from Eqs.~(\ref{psym1}) and (\ref{psym2}) that the perturbed CVs are antisymmetric
\begin{eqnarray}
\left(\frac{\partial L_{\sigma \rho}^P}{\partial \bf B}\right)_{{\bf B}= {\bf 0}} = - \left(\frac{\partial L_{\rho \sigma}^P}{\partial \bf B}\right)_{{\bf B}= {\bf 0}}
\end{eqnarray}
in agreement with the discussion in Ref.~\onlinecite{Burger21}.

The first-order Taylor expansion also provides an expression for the perturbed CVs. We obtain here for the real case
\begin{eqnarray}
\frac{\partial M_{\sigma \rho}^{P\rightarrow (\mu \nu)}}{\partial {\bf B}} = \widetilde{(\mu \nu | \nu \mu)}^{- \frac{1}{2}}
\left\{\frac{\partial(\sigma \rho | \nu \mu )}{\partial {\bf B}} - \sum_{Q=1}^{P-1} \left(\frac{\partial M_{\sigma \rho}^Q}{\partial {\bf B}} {M_{\mu\nu}^Q}^* +  M_{\sigma \rho}^Q \frac{\partial{M_{\mu\nu}^Q}^*}{\partial {\bf B}}\right) \right\} 
\end{eqnarray}
and for the complex case for the perturbed CVs in the partially orthogonalized representation
\begin{eqnarray}
\frac{\partial \overline{M}_{\sigma \rho}^{P \rightarrow (\mu \nu)}}{\partial {\bf B}} = 
\frac{\partial(\sigma \rho | \nu \mu )}{\partial {\bf B}} - \sum_{Q=1}^{P-1} \left( \frac{\partial M_{\sigma \rho}^Q}{\partial {\bf B}} {M_{\mu\nu}^Q}^* + M_{\sigma \rho}^Q \frac{\partial {M_{\mu\nu}^Q}^*}{\partial {\bf B}}
\right)
\end{eqnarray}
\begin{eqnarray}
\frac{\partial \overline{M}_{\sigma \rho}^{P^\prime \rightarrow (\nu \mu)}}{\partial {\bf B}} &=&
\frac{\partial(\sigma \rho | \mu \nu )}{\partial {\bf B}} - \sum_{Q=1}^{P-1} \left( \frac{\partial M_{\sigma \rho}^Q}{\partial {\bf B}} {M_{\nu\mu}^Q}^* + M_{\sigma \rho}^Q \frac{\partial {M_{\nu\mu}^Q}^*}{\partial {\bf B}}
\right)
. 
\end{eqnarray}
Using Eqs.~(\ref{eq:derlim1}) and (\ref{eq:derlim2}), the latter in its symmetric version given in Eq.~(\ref{pertlc}), we get in the zero field limit the following expression for the perturbed CVs
\begin{eqnarray}
\label{intder2}
\frac{\partial L_{\sigma \rho}^{P\rightarrow (\mu \nu)}}{\partial {\bf B}} = \widetilde{(\mu \nu | \nu \mu)}^{- \frac{1}{2}}
\left\{\frac{1}{2}\left\{\frac{\partial (\sigma \rho | \nu \mu )}{\partial {\bf B}} + \frac{\partial (\sigma \rho | \mu \nu )}{\partial {\bf B}}\right\} - \sum_{Q=1}^{P-1} \frac{\partial L_{\sigma \rho}^Q}{\partial {\bf B}} {L_{\mu\nu}^Q}^*\right\}
\end{eqnarray}
Note that Eq.~(\ref{intder2}) holds for both the real and complex case.

The expression in Eq.~(\ref{intder2}) for the perturbed CVs can be further rewritten in terms of partially differentiated two-electron integrals
\begin{eqnarray}
\label{intder3}
\frac{\partial L_{\sigma \rho}^{P\rightarrow (\mu \nu)}}{\partial {\bf B}} = \widetilde{(\mu \nu | \nu \mu)}^{- \frac{1}{2}}
\left\{ (\frac{\partial \sigma \rho }{\partial {\bf B}}| \nu \mu )  - \sum_{Q=1}^{P-1} \frac{\partial L_{\sigma \rho}^Q}{\partial {\bf B}} {L_{\mu\nu}^Q}^*\right\},
\end{eqnarray}
thereby exploiting that
\begin{eqnarray}
\frac{\partial \sigma \rho| \nu \mu)}{\partial \bf B} &=& (\frac{\partial \sigma \rho}{\partial \bf B} |\mu \nu ) +
(\sigma \rho |\frac{\partial \nu \mu}{\partial \bf B} )\nonumber \\ &=&
 (\frac{\partial \sigma \rho}{\partial \bf B} |\mu \nu ) -
(\sigma \rho |\frac{\partial \mu \nu}{\partial \bf B} ).
\end{eqnarray}
Use of Eq.~(\ref{intder3}) instead of Eq.~(\ref{intder2}) means that the cost for the integral evaluation in the CD procedure for magnetic two-electron integral derivatives can be reduced by a factor of four, as (a) only one integral derivative is needed instead of two and as (b) the cost for computing this partial derivative is only half of those for the full integral derivative. 
In a further step, one can compute the magnetic two-electron integral derivatives within the already discussed two-step CD procedure. Here, we use the observation that the CD for the magnetic two-electron integral derivatives corresponds to density fitting with the unperturbed Cholesky basis. One can thus apply the two-step CD procedure such that only in the second step the standard integrals are replaced by their partial derivatives, i.e., the perturbed CVs are obtained as
\begin{eqnarray}
\frac{\partial L_{\sigma \rho}^{P\rightarrow (\mu \nu)}}{\partial {\bf B}} = \sum_Q (\frac{\partial \sigma \rho }{\partial {\bf B}}| Q ) (Q|P)^{-1/2}
\end{eqnarray}
This means that the same computational savings as for the standard two-electron integrals can be realized for the magnetic two-electron integral derivatives by using the two-step CD procedure.

\section{Results}

The CD procedures described in the previous section have been implemented within the CFOUR program package.\cite{cfour,Matthews20c} In Table~\ref{table1} we document the computational advantages of our modified CD procedure for finite magnetic-field calculations. To be more specific, we compare the performance of our new CD procedure with those of the original procedure from Ref.~\onlinecite{Blaschke22} with respect to the number of needed CVs and the required memory to store the CVs. In addition, we analyze in detail how many CVs correspond to real and how many correspond to complex products and investigate the performance of our new CD procedure for different magnetic field strengths. Results are reported in Table~\ref{table1} for calculations on ethane (CCSD/cc-pVTZ geometry for ${\bf B} = {\bf 0}$) with magnetic field strengths of 0.1, 0.5, and 1.0 a.u. (1 a.u. for the magnetic flux density corresponds to about 235052 Tesla)
with the magnetic field oriented along the CC bond. As basis set, we use the cc-pVTZ set (144 basis functions) from Dunning' hierarchy of correlation-consistent basis sets.\cite{Dunning89} As examples for the benefits of the new procedure for calculations on larger molecules, we present results for the corannulene dianion and the retinal molecule using the uncontracted cc-pVDZ basis\cite{Dunning89} (unc-cc-pVDZ, 590 and 742 basis functions, respectively) and a magnetic field of 0.1 a.u. The geometries used in these calculations are the same as the ones already used in Ref.~\onlinecite{Blaschke22}.

From these calculations, we note that the new CD procedure leads to a slightly increased number of CVs which is expected, as for each complex product, unlike in the original CD procedure, the conjugate product is always added to the Cholesky basis. However, the increase is modest and in all investigated cases amounts to only a few percent. The analysis of the character of the considered products for the Cholesky basis reveals that interestingly a significant amount of products or even the majority of products are real. For the corannulene dianion more than 75 \% of the products are real and we also note that the ratio of real to complex products is, as expected, also to a large part independent of the strength and orientation of the applied magnetic field. From a computational perspective, most important are the memory requirements. Here we see that the new CD procedure requires in comparison to the original scheme only about half of the memory to store all CVs compared to the old CD scheme. In case of the corannulene dianion this means that instead of about 16 GB the new CD algorithm only needs slightly more than 8 GB for storage of the CVs. This can be considered a significant advantage, as for the actual computations it is most advantageous when the full set of CVs can be kept in memory, and these savings hold for the CVs as computed in the atomic-orbital basis as well as for the CVs transformed into the molecular-orbital representation, that are preferably used in electron-correlated calculations. At this point we also note that the exploitation of the symmetry of the CVs also offers the potential for savings in the actual computational timings, but those have not been explored in the present work.
\begingroup
\begin{table}
\caption{Comparison of original and new CD procedure concerning number of CVs ($N_{\rm CV}$) and the memory (in MB) required to store the full set of CVs. The comparison is made for the ethane molecule (C$_2$H$_6$), thereby considering different magnetic-field strengths (in a.u.), as well as the corannulene dianion (C$_{20}$H$_{10}^{2-}$) and the retinal molecule (C$_{20}$H$_{28}$O) and includes for the new CD procedure information about the number of real and complex products in the Cholesky basis. Calculations for the ethane molecule were carried out with a Cholesky threshold of 5, while the other calculations used a Cholesky threshold of 4.}
    \label{table1}
    \centering
    \begin{tabular}{cccccccccc}
        \hline
         & & & & \multicolumn{2}{c}{original CD}&\quad& \multicolumn{3}{c}{new CD} \\ \cline{5-6}\cline{8-10}
        molecule &\quad & basis ($N_{bf}$) & magnetic field& $N_{CVs}$ & memory &\quad& $N_{\rm CVs}$& $N_{\rm CVs}$(real)  & memory \\
       \hline
       C$_2$H$_6$ && cc-pVTZ (144) & 0.1 & 1150 & 364 && 1213 & 663 (  54.5\%) & 179 \\
       & & & 0.5 & 1250 & 396 && 1284 & 640 (  49.8\%) & 204\\
       & & & 1.0 & 1354 & 429 && 1360 & 610 (  44.9\%) & 216\\
       C$_{20}$H$_{10}^{2-}$ & \ \ \quad\  & unc-cc-pVDZ (590) & 0.1(perpendicular) & 2938 & 15606 &\ \quad\  & 3077 & 2425 (78.8\%) & 8183\\
       & \ \ \quad\  &  & 0.1 (parallel) & 2991 & 15887 &\ \quad\  & 3147 & 2389 (75.9\%) & 8369 \\ 
        C$_{20}$H$_{28}$O & \ \ \quad\  & unc-cc-pVDZ (742) & 0.1 & 3636 & 30546 &\ \quad\  & 3878 & 2946 (76.0\%) & 16307 
              \\
       \hline
\end{tabular}
\end{table}
\endgroup

Concerning the CD of the magnetic integral derivatives, a comparison of the computational timings for the original, the improved, as well as the two-step algorithm can be found in Table~\ref{table2}. Timings are given for the molecules that 
were already used as examples in Ref.~\onlinecite{Burger21}, i.e.,
coronene (C$_{24}$H$_{12}$), hexabenzocoronene (C$_{42}$H$_{18}$), tetrakis(t-butyl)tetraborane(4) (B$_4$C$_{16}$H$_{36}$), tetrameric cyclopentadienyl aluminum(I) (Al$_4$C$_{20}$H$_{20}$), and the buckminsterfullerene (C$_{60}$). The structures of these molecules have been depicted in Figure 2 of Ref.~\onlinecite{Burger21} and the corresponding Cartesian coordinates have been given in the supplementary material of the same reference. Calculations 
have been carried out with the dzp and tz2p basis sets\cite{Schaefer92,Gauss93} that also have been used in Ref.~\onlinecite{Burger21} and have been documented in the corresponding supplementary material. We report in Table~\ref{table2} timings for the integral evaluation as well for the construction of the actual CVs for the standard two-electron integrals and the magnetic integral derivatives. In this way we document the improvements in the integral evaluation due to the use of the partial instead of the full derivatives and the savings in the construction of the CVs due to the use of a two-step procedure. All calculations reported in Table~\ref{table2} were carried out with a Cholesky threshold of 5.
\begingroup
\begin{table}
\scriptsize
\caption{Comparison of the timings (in seconds) for the CD of standard two-electron integrals and magnetic two-electron integral  derivatives using the original algorithm from Ref.~\onlinecite{Burger21}, the improved algorithm using partial integral derivatives, and the two-step procedure from Ref.~\onlinecite{Folkestad19}. $t_{\rm Int}$, $t_{\rm CV}$, and $t_{\rm CD}$
denote the timings for integral evaluation, for the construction of the CVs, and for the total CD procedure, respectively, while $f_{\rm Int}$, $f_{\rm CV}$, and $f_{\rm CD}$ give the corresponding speedups compared to the original CD procedure. All calculations have been carried out using one core of an Intel Xeon(R) Gold 6342 node running at 2.8 GHz.}
    \label{table2}
    \centering
    \begin{tabular}{lcccccccccccccccc}
        \hline
         & & &&  \multicolumn{6}{c}{standard 2el integrals}&\quad& \multicolumn{6}{c}{magnetic 2el integral derivatives} \\ \cline{5-10}\cline{12-17}
        molecule  & basis  & $N_{bf}$ & algorithm & $t_{\rm Int}$ & $t_{\rm CV}$ & $t_{\rm CD}$& $ f_{\rm Int}$& $f_{\rm CV}$& $f_{\rm CD}$ &\quad& $ t_{\rm Int}$ & $t_{\rm CV}$ & $t_{\rm CD}$ & $ f_{\rm Int}$& $f_{\rm CV}$& $f_{\rm CD}$ \\
       \hline
       C$_{24}$H$_{12}$ & dzp &420& original & 123 & 79 & 204 &&&&& 1111 & 241 & 1360 \\
                &       && new & &  &  &&&&& 245 & 241 & 494 & 4.5 & 1.0 & 2.8  \\
                &       && two-step & 181 &  8 & 85 & 0.7 & 9.6 & 2.4 && 235 & 25 & 262 & 4.7 & 9.7 & 5.2\\
                       C$_{24}$H$_{12}$ & tz2p &684& original & 401 & 507 & 908 &&&&& 4161 & 1534 & 5695 \\
                &       && new &  &  &  &&&&&942 & 1534& 2476 & 4.4 & 1.0 & 2.3 \\
                &       && two-step & 632 & 57 & 393 & 0.6 & 8.8 & 2.3 && 1009 & 166 & 1183 & 4.1 & 9.2 & 4.8 \\
                       C$_{42}$H$_{18}$ & dzp &720& original & 420 & 702 & 1129 &&&&& 3875  & 2121 & 6034 \\
                &       && new &  &  &  &&&&& 866 & 2119 & 3021 & 4.5 & 1.0 & 2.0 \\
                &       && two-step & 597 & 71 & 713 & 0.7 & 9.9 & 1.6 && 845 & 212 & 1066  & 4.6 & 10.00 & 5.7 \\
                       C$_{42}$H$_{18}$ & tz2p &1170& original & 1414 & 4361 & 5809 &&&&& 14602 & 13158 & 27943 \\
                &       && new &  &  &  &&&&& 3338 & 13251 & 16760& 4.4 & 1.0 & 1.7 \\
                &       && two-step & 2214 & 473 & 3015 &0.6 &9.2 & 1.9&& 3675 & 1422 & 5159 & 4.0 & 9.3 &  5.4 \\
                                       B$_4$C$_{16}$H$_{36}$ & dzp &480& original & 154 & 133 & 289 &&&&& 1264 & 403 & 1678 \\
                &       && new &  &  &  &&&&& 292 & 403 & 705 & 4.3 & 1.0 & 2.4 \\
                &       && two-step & 231 & 14 & 125 & 0.7 & 9.5 & 2.3 && 291 & 43 & 336 & 4.3 & 9.4 & 5.0 \\
                       B$_4$C$_{16}$H$_{36}$ & tz2p & 804& original & 524 & 876 & 1410 &&&&& 4865 & 2623 & 7541 \\
                &       && new &  &  &  &&&&& 1117 & 2622 & 3789 & 4.4 & 1.0 & 2.0 \\
                &       && two-step & 824 & 91 & 567 & 0.6 & 9.6 & 2.5 && 1265 & 272 & 1551 & 3.8 & 9.6 & 4.8 \\                       Al$_4$C$_{20}$H$_{20}$ & dzp &492& original & 146 & 139 & 288 &&&&& 1321 & 421 & 1755 \\
                &       && new &  &  &  &&&&& 296 & 421 & 729 & 4.5 & 1.0 & 2.4 \\
                &       && two-step & 211 & 14 & 243 & 0.7 & 9.9 & 1.2 && 291 & 43 & 336 & 4.5 & 9.8 & 5.2 \\
                       Al$_4$C$_{20}$H$_{20}$ & tz2p &788& original & 461 & 842 & 1313 &&&&& 4817 & 2543 & 7413 \\
                &       && new &  &  &  &&&&& 1078 & 2541 & 3669 & 4.5 & 1.0 & 2.0\\
                &       && two-step & 728 & 89 & 911 &0.6 &9.5 & 1.4 && 1188 & 273 & 1474 & 4.1 & 9.3 & 5.0 \\                       C$_{60}$ & dzp &900& original & 1047 & 1682 & 2744 &&&&& 10061 & 5081 & 15220 \\
                &       && new &  &  &  &&&&& 2226 & 5076 & 7376 & 4.5 & 1.0 & 2.1 \\
                &       && two-step & 1402 & 180 & 1356 &0.7 &9.3 & 2.0 && 2137 & 543 & 2653 & 4.7 & 9.4 & 5.7 \\
                       C$_{60}$ & tz2p &1440& original & 3529 & 10832 & 14427 &&&&& 37203 & 32175 & 69741 \\
                &       && new &  &  &  &&&&& 8346 & 32156 & 40841 & 4.5 & 1.0 & 1.7\\
                &       && two-step & 5648 & 1129 & 7968 &0.6 & 9.6 & 1.8 && 9785 & 3765 & 13804 & 3.8 & 8.5 & 5.1 \\
          \hline
\end{tabular}
\end{table}
\endgroup
The timings reported in Table~\ref{table2} show that the savings in the evaluation of the magnetic two-electron integral derivatives due to the use of the partial instead of the full derivatives are significant. The expected speed-up was about four; the timings show even larger speed-ups. For example, for the tz2p calculation on hexabenzocoronene with 1170 basis functions, the old CD algorithm required about 14600 seconds for the evaluation of the magnetic two-electron integral derivatives, while the new one only needs about 3340 seconds which corresponds to a speed-up of about 4.4. However, the speed-up for the whole CD procedure is lower, e.g. for the just mentioned hexabenzocoronene case we observe here only a speed-up of 1.7. This finding is explained by the fact that within the original CD procedure a large amount of the time is spent for the construction of the actual CVs. This computational bottleneck is tackled with our second improvement, i.e., the use of a two-step CD procedure as proposed in Refs.~\onlinecite{Folkestad19,Zhang21}. The use of a two-step procedure leads to significantly lowering of the computational cost for the CD of the standard integrals as well as the magnetic integral derivatives. While the savings in the timings for the construction of the CVs are substantial (about a factor of 9 to 10 for the standard integrals as well as the integral derivatives), the savings are somewhat reduced due to the increased cost for the integral evaluation. This increase is due to the fact that within the two-step procedure the integrals have to be computed for the selection of the Cholesky basis as well as for the construction of the CVs. The overall savings for the CD of the standard two-electron integrals thus are less pronounced, though one still observes speed-ups in the range of 1.5 and 2.5.
For the magnetic integral derivatives, however, no additional overhead in the integral evaluation occurs, as the same Cholesky basis as for the standard two-electron integrals is used and the savings in the construction of the CVs are fully realized. The speed-up for the hexabenzocoronene case, for example, is about 5.4 and thus is significant. The other cases listed in Table~\ref{table2} support these findings with speed-ups around 5.
Note that we ran all calculations on purpose on only one core, as we wanted to exclude side effects due to varying parallelization efficiencies in the different steps. It is clear that the CD algorithms used can be significantly
accelerated by parallelizing both the integral evaluation as well as the construction of the CVs. For the latter,
an easy way to exploit parallelization is to use a multithreaded blas or lapack library, as has been already done
in Ref.~\onlinecite{Burger21}. We also note that the reported timings can be further reduced by additional optimization of our codes and that in our CD procedures sparsity has not been exploited, but it is clear that this might be of advantage for large-scale cases as discussed, for example, in Ref.~\onlinecite{Folkestad19}.

\section{Conclusions and outlook}

In this article, we present a detailed theoretical and numerical analysis concerning the CD of two-electron integrals in case of quantum-chemical calculations with finite or perturbative magnetic fields. For the finite magnetic-field case we show how the fourfold permutational symmetry of the two-electron integrals that are decomposed can be accounted for and exploited. We propose a modified CD procedure which essentially requires to include for those
Cholesky basis functions that correspond to complex products of the original basis functions also the corresponding complex conjugate function. This modified CD procedure does not only preserve the symmetry but also offers the potential for computational savings. Noteworthy and documented in the present paper are the savings of about a factor of two in the memory requirements for the storage of the full set of Cholesky vectors.

For the magnetic two-electron integral derivatives, we deduce the antisymmetry of the corresponding perturbed CVs
by means of a first-order Taylor expansion for the corresponding finite-field expressions around the field-free case as reference point. We suggest that computational savings are possible (a) by the use of partial instead of full integral derivatives in the integral evaluation step, and (b) by means of a two-step procedure to speed up the actual construction of the unperturbed and perturbed CVs from the computed integral derivatives. These savings in the computational timings are verified in sample calculations where we note savings of more than four for the integral evaluation step and of more than nine in the construction of the CVs. The total speed-up in the construction of the perturbed CVs amounts to more than five and is in particular relevant for computations, e.g., at the CD-CASSCF level,\cite{Nottoli22b} where the CD step constitutes a computationally significant step.

Future work will focus on exploiting the symmetry in the CVs for finite magnetic-field calculations at the coupled-cluster level and on the extension of the presented discussion to magnetic integral derivatives for corresponding second derivatives that are needed for the computation of magnetizabilities.\cite{Ruud93,Gauss07}

\begin{acknowledgments}
This paper is dedicated to Professor Péter G. Szalay on the occasion of his 60th birthday. One of the authors (J.G.) thanks Péter G. Szalay for more than thirty years of friendship, the hospitality during many visits to Budapest, and
collaboration in numerous scientific projects.
The authors acknowledge funding by the Deutsche Forschungsgemeinschaft (DFG) within project B5 of the TRR 146 (project no. 233 630 050). S.S. also acknowledges support from the DFG via grant STO 1239/1-1.
\end{acknowledgments}
\label{sec:cc}

\bibliography{cdgiaomp2}

\end{document}